\documentclass[%
aip,
amsmath,amssymb,
reprint,%
]{revtex4-1}
\usepackage{graphicx}
\usepackage{dcolumn}
\usepackage{bm}
\usepackage[utf8]{inputenc}
\usepackage[T1]{fontenc}
\usepackage{mathptmx}
\usepackage{etoolbox}
\usepackage{subcaption}
\usepackage[colorlinks=false,linkcolor=blue]{hyperref}
\usepackage{xcolor, soul} 
\makeatletter
\def\@email#1#2{%
	\endgroup
	\patchcmd{\titleblock@produce}
	{\frontmatter@RRAPformat}
	{\frontmatter@RRAPformat{\produce@RRAP{*#1\href{mailto:#2}{#2}}}\frontmatter@RRAPformat}
	{}{}
}%
\makeatother
\begin{document}
	\preprint{AIP/123-QED}
	\title{Controlling extreme events in neuronal networks: A single driving signal approach}
	\author{R. Shashangan}
	\email{shashang.physics@gmail.com}
	\affiliation{Department of Nonlinear Dynamics, Bharathidasan
		University, Tiruchirappalli 620024, Tamil Nadu, India.}
	\author{S. Sudharsan}
	\email{dr.sudharsan32@gmail.com}
	\affiliation{Center for Nonlinear and Complex Networks, SRM TRP Engineering College, Tiruchirapalli 621105, Tamil Nadu, India.}
	\affiliation{Center for Research, Easwari Engineering College, Chennai 600089, Tamil Nadu, India.}
	\author{Dibakar Ghosh}
	\email{diba.ghosh@gmail.com}
	\affiliation{Physics and Applied Mathematics Unit, Indian Statistical Institute, 203 B. T. Road, Kolkata 700108, India.}
	\author{M. Senthilvelan$^{1}$}
	\email{velan@cnld.bdu.ac.in (Corresponding Author)}
	\begin{abstract}
		We show that in a drive-response coupling framework extreme events are suppressed in the response system by the dominance of a single driving signal. We validate this approach across three distinct response network topologies, namely (i) a pair of coupled neurons, (ii) a monolayer network of N coupled neurons and (iii) a two-layer multiplex network each composed of FitzHugh-Nagumo neuronal units. The response networks inherently exhibit extreme events. Our results demonstrate that influencing just one neuron in the response network with an appropriately tuned driving signal is sufficient to control extreme events across all three configurations.  
		In the two-neuron case, suppression of extreme events occurs due to the breaking of phase-locking between the driving neuron and the targeted response neuron. In the case of monolayer and multiplex networks, suppression of extreme events results from the disruption of protoevent frequency dynamics and a subsequent frequency decoupling of the driven neuron from the rest of the network. We also observe that when the size of the neurons in response network connected to the drive increases, the onset of control occurs earlier indicating a scaling advantage of the method.
	\end{abstract}
	\maketitle
	\begin{quotation}
		\textbf{Intermittent, unanticipated, and sudden synchronization of nodes consisting of neuronal oscillators manifest a phenomenon called extreme events. This phenomenon occurring in a mathematical framework can be correlated with an important physiological condition, the epileptic seizure, which occurs as a result of sudden synchronization of the firing of neurons inside the brain. These seizures are a serious threat to humans and animals as they are possibly fatal. Understanding their dynamics through the lens of the mathematical framework of networks provides us the space to comprehend the properties of these events, such as origin, propagation, sustenance and mitigation, more deeply. Formulating a strategy and devising tools to control these undesirable events is an important task. The present study fundamentally addresses a way of controlling these events in a network. The procedure proposed in this paper is an efficient one that deploys a drive system exhibiting relaxation oscillations to interact and control the extreme events in the response networks upon influencing. This drive-response technique to control extreme events is demonstrated on (i) a simple two-coupled system, (ii) a monolayer $N$-coupled network, and (iii) a two-layer multiplex network. The usefulness of this study lies in the ease of realization of the methodology.}
	\end{quotation}
	\section{Introduction}
	\label{Intro}
	\par Extreme events (EEs) are rare and recurrent catastrophic events that appear in a variety of natural and man-made systems \cite{Kantz1, Najafi1}. The study of EEs in dynamical systems encompasses the determination of the (i) mechanism, (ii) statistical behaviour, (iii) prediction tools, and (iv) mitigation strategies \cite{Farazmand}. The fundamental dynamical behaviour of a range of EEs, such as tsunamis, earthquakes, epileptic seizures, epidemics, share market crashes, and power blackouts, can be understood only by investigating these phenomena in their corresponding dynamical systems. {\textcolor{black}{The occurence of epileptic seizures is one of the visible applications of EEs in biological context. In this perspective, EEs have been observed in the region of onset of epileptic-seizure-related (ESR) sychronization regimes in small-world networks of Fitzhugh-Nagumo (FHN) oscillators \cite{Cubillos}. ESR sychronization patterns have also been observed in FHN network(s) under different network topologies. These patterns replicate the patterns of electroencephalographic (EEG)-recorded epileptic seizures \cite{Gerster}. We note that in the literature extreme value theory (EVT) has also been used to classify epileptic seizures. For instance, EVT has been used to study the statistical nature of seizures in the EEG data of stroke-induced epilepsy in mouse \cite{Frolov1} and in the EEG data of epileptic seizures in WAG/Rij rats to understand its statistical properties and to predict the onset of seizures \cite{Frolov2}. It is important to note that epileptic seizures are manifested as a sudden, abnormal, and excessive synchronization among neurons in the brain \cite{Kantz2}. Interestingly, such a phenomenon has also been observed even in a network of neurons whose time evolutions are represented by a set of differential equations \cite{Izhikevich, Gerstner}. Examining the dynamics of network of oscillators facilitate a simple but effective analysis of this phenomena theoretically.}}
	\par Analyzing them on a smaller scale is one of the easier ways to understand their dynamics, source of emergence, way of propagation, and so on. EEs have been extensively studied in isolated dynamical systems \cite{Review}. In the context of some applications pertaining to networks, to understand the functioning of EEs on a large scale, it is essential to study them in coupled systems/networks. {\textcolor{black}{This is because networks are the best way to represent and understand the structure and functioning of various complex systems, as they provide a complete description of how the individual units are connected and how the individual units and the system as a whole evolve dynamically. Further, they help to uncover the underlying mechanism not only on the properties of complex systems but also changes occurring due to external perturbation,  various collective phenomena emerging in them \cite{Gosak} and so on. In addition, the nodes, links or central hubs which are vulnerable to attacks and cause system failure due to external perturbations can be identified easily using the measures of networks. Network science plays a crucial role in making the system highly resilient and robust \cite{Artime}.}}  In this connection, properties of EEs in networks have been observed in the coupled and network of FHN neuron model \cite{Ansmann}, coupled Hindmarsh-Rose (HR) neuronal model \cite{Mishra1, SS3}, network of Josephson junctions \cite{Mishra2}, coupled R\"{o}ssler system \cite{SS4}, small-world networks \cite{Ansmann}, network of neuronal and chaotic maps \cite{SNC, SSinha3}, population models \cite{Yukalov, SSinha1, SSinha2}, micro-electro-mechanical system (MEMS) cantilevers \cite{Pisarchik2}, in moving agents \cite{Perc1}, minimal universal laser (MUL) network model \cite{Perc2}, and Bak–Tang–Wiesenfeld (BTW) model \cite{Najafi2}. Similarly, EEs have also been explored in multilayer networks \cite{Seager, Shashang2}.
	\par Though preventing natural EEs is strenuous, controlling them in applied systems is significant. It helps to prevent damage imminent to the machinery and to ensure their smooth operation \cite{Review, Shrimali1, Shrimali2}. Dynamical systems come in handy to help to create strategies towards control aspects. A well modelled mitigation technique using dynamical systems can help to explore, extend, and apply them to large-scale systems. In both isolated and coupled dynamical systems, various tools such as constant bias \cite{Shashang1, SS1}, external forcing \cite{SS1}, threshold activated coupling \cite{Aray}, time-delayed feedback \cite{SS2, Suresh}, noise \cite{Zamora}, localized perturbation \cite{Bialonski}, network mobiling \cite{Chen} and environment coupling \cite{Varshney} have been used to mitigate EEs. Similarly, synchronous firing of neurons is controlled using channel blocking \cite{Uzuntarla1} and astrocytes\cite{Uzuntarla2}. Effective to work, versatile to apply, and easy to implement are some of the prerequisites that necessitate the back to back creation of novel control methods for EEs. 
	\begin{figure}[!ht]
		\centering
		\includegraphics[width=1.0\linewidth]{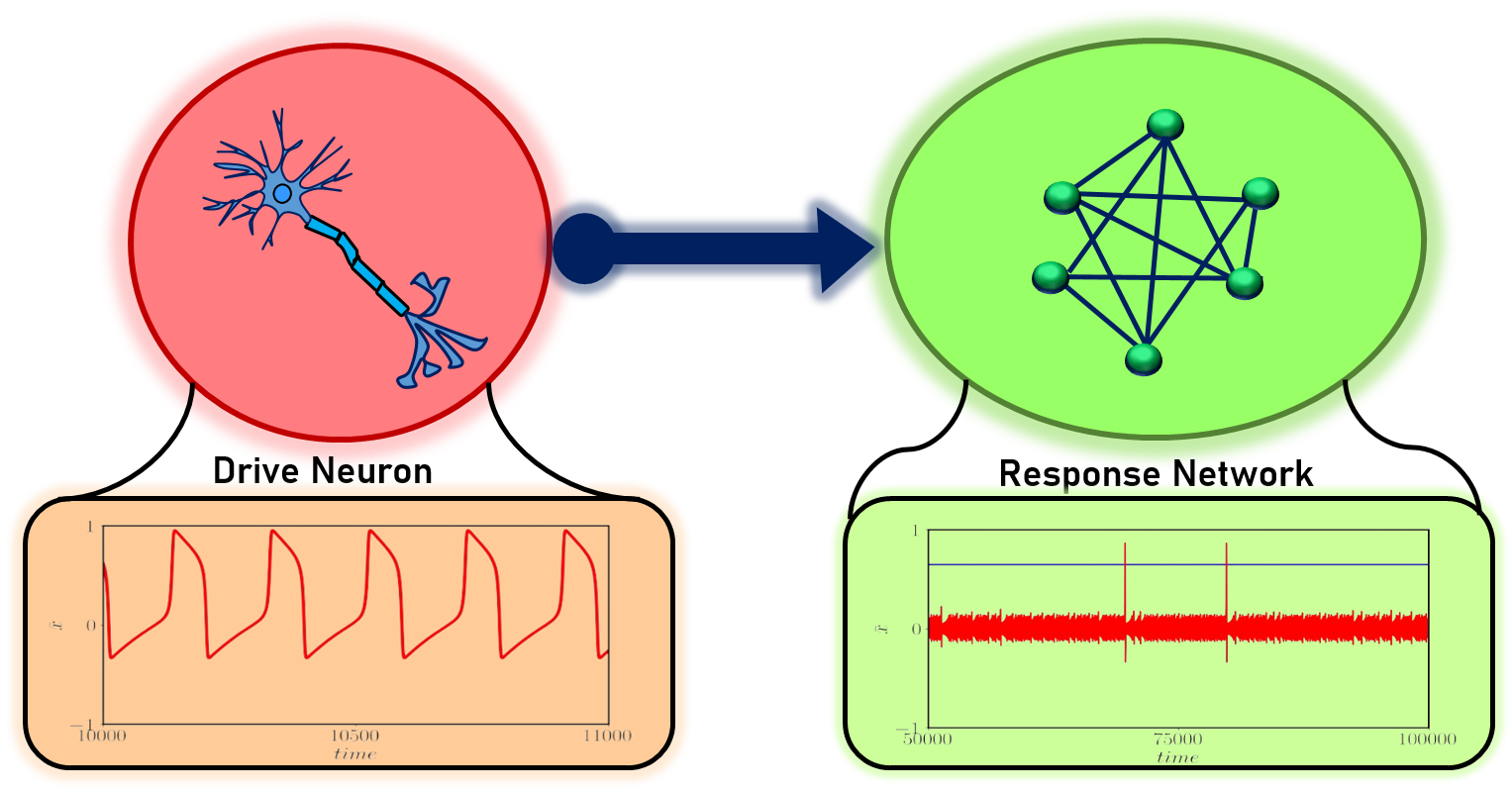}	
		\caption{Drive-response coupling configuration where a single neuron in the drive (red circle) and a neuronal network in the response system (green oval). The drive neuron exhibits {\textcolor{black}{spiking in the form of relaxation oscillations}} (red box) while the response network as a whole exhibits chaotic dynamics with EE (green box).}
		\label{gen_config}
	\end{figure}
	\par In this work, we consider a drive-response coupling configuration, well-known in electronics, to eliminate EEs appearing in response networks. Particularly, we make use of the fact that ``drive influences the response" and use this as an approach to eliminate EEs. Notwithstanding the elimination of EEs, we also determine the conditions and the scientific reasoning behind the suppression of EEs. The drive-response coupling configuration that we consider in this manuscript consists of only one neuron in the drive and a network of neurons in the response. The single drive neuron drives the response network throughout the time. Further, the coupling strength between the drive and the response determines the magnitude by which the drive influences the dynamics of the response. The usage of drive-response approach is ubiquitous. For example, it is used to improve the torque performance of the motor \cite{torque}, to synchronize the motors in ocean winch \cite{ocean winch}, in magnetic resonance imaging (MRI) for guided interventions \cite{MRI}, to prevent the radiation leakage in nuclear industries \cite{radiation}, to perform the single-port laparoscopy (SPL) surgery \cite{SPL}, to achieve synchronization for secure communications \cite{SC1, SC2}, for hierarchial, power and voltage control in microgrid systems \cite{HC, PVC} and in machine learning based harvest control in agriculture \cite{Harvest}. We adopt this strategy of influence and investigate the elimination of EEs arising from a neuronal network. A schematic representation of this idea is illustrated in Fig.~\ref{gen_config}. In this figure, the red circle on the left hand side represents a neuron for the drive; the corresponding dynamics is presented in the red box. The green oval on the right hand side in Fig.~\ref{gen_config} denotes the response network, and its corresponding dynamics is presented in the green box. In between, the blue arrow pointing towards right, denotes that information can flow only from drive to response. The dynamics exhibited by the drive neuron are {\textcolor{black}{spiking in the form of relaxation oscillations}} (typical regular dynamics of the FHN neuron model). The importance of choosing relaxation oscillations in the drive is that (i) it operates in the periodic regime, and (ii) it can be generated easily in laboratories. The response neuron network is configured in a manner that the mean field of the membrane potential of the network exhibits chaotic dynamics, accompanied by EEs (see the green box in Fig.~\ref{gen_config}). We will also demonstrate that a drive neuron connected to only one neuron in the response network is more than sufficient to make these events disappear in the response network.
	\par This approach is validated across three distinct response network topologies: (i) a pair of coupled neurons, (ii) a monolayer network of $N$ coupled neurons, and (iii) a two-layer multiplex network, each comprising FitzHugh-Nagumo neuronal units. {\textcolor{black}{Response network (i) is chosen to understand fundamentally the dynamics behind the control from the perspective of neuron to neuron interaction, while the rationale behind the choice of response networks (ii) and (iii) stems from the necessity to explore control efficacy under increased complexity and model the functioning of interconnected brain regions through interlayer coupling. Further the necessity of studying the response networks (ii) and (iii) is to understand the two broad types of epileptic seizures, namely (i) focal - where the epilepsy are localized in particular regions and (ii) generalized - where the onset of epilepsy begins in a particular area but can spread across regions and hemispheres \cite{Fisher}. Through our analysis, we find that}} in the case of (i) mitigation occurs through breaking of phase-locking whereas in the case of (ii) $\&$ (iii) it occurs through the disruption of protoevents' frequency. Furthermore, we observe that, in all three cases, as the coupling strength  between the drive and response neuron is increased, the frequency of the extreme events decreases gradually and they are abruptly suppressed. We also find an interesting fact that when we increase the number of response neurons connected to the drive the onset of control occurs earlier.
	\par Phenomenologically, drive-response coupling configuration is well known for stimulating various types of synchronization, such as generalized synchronization \cite{Rulkov}, complete synchronization \cite{Buscarino}, and partial synchronization \cite{Legorreta}. Further, it has also been used for the estimation of synchronization parameters \cite{Joaquin}, and even in the promotion of EEs \cite{Hugo}. Differing from the above, here, the drive-response configuration destroys the synchronization happening among the neurons in the system. Specifically, we disrupt the unanticipated, intermittent complete synchronization happening occasionally among neurons in different networks of FHN neurons. The advantage of using drive-response coupling to eliminate EEs in neuronal networks is that this approach can be easily extended even to real-time brain studies. Since it is easy to generate relaxation oscillations using any commercial oscilloscope, and apply the generated signal to the brain of the patient showing signs of epilepsy. 
	
	\par We organize our study as follows. In Sec.~\ref {mad}, we define the mathematical model of the drive-response network under consideration and the measure used for our analysis. In Sec.~\ref{mitigation}, we discuss the mitigation of EEs in three different FHN networks. The mechanism behind controlling EEs is discussed in Sec.~\ref{mecha}. In Sec.~\ref{pshift}, we discuss the relation between the coupling strength and number of response neurons. Finally, we conclude our results in Sec.~\ref{conc}. 
	\section{Proposed models} 
	\label{mad}
	We start our analysis by considering a drive-response coupling configuration where the drive influences the response throughout. We represent the generalized mathematical model of the considered drive neuron and the response network in the form
	\begin{eqnarray}
		&\dot{\textbf{X}}&=\textbf{F}(\textbf{X}), \nonumber\\
		&\dot{\textbf{Y}}_i&=\textbf{G}(\textbf{Y}_i)+k\sum_{j=1}^N A_{ij}~\textbf{g}(\textbf{X},\textbf{Y}_1),
	\end{eqnarray}
	where $\dot{\textbf{X}}$ and $\dot{\textbf{Y}}_i$ are vectors consisting of the state variables representing the time evolution of the drive neuron and response neurons network, respectively. Further, $\textbf{g}(\textbf{X},\textbf{Y}_1)$ is the coupling function that couples the drive neuron with the first neuron in the response system. In this coupling configuration, we observe that for even a very weak coupling between the drive and response neurons, the EEs in the response network are mitigated.  As discussed in the introduction section, a single neuron is utilised as a drive. In the place of the response, three kinds of neuronal networks are considered, namely (i) two-coupled, (ii) monolayer N-coupled and (iii) two-layered multiplex networks are taken and tested for suppression. Isolated Fitzhugh-Nagumo neuron model operating in the relaxation oscillation phase is taken as the dynamical equation for the drive. The corresponding dynamical equation is \cite{Izhikevich}
	\begin{eqnarray}
		\dot{x}_m &=& x_m(a - x_m)(x_m - 1) - y_m + I, \nonumber \\
		\dot{y}_m &=& bx_m - cy_m.  
		\label{fhn_mas}
	\end{eqnarray}
	Here, the suffix $m$ denotes the drive neuron. $x$ and $y$ represent the membrane potential and recovery variable of a neuron, respectively. The parameter $a$ describes the shape of the cubic polynomial, while $b$ and $c$ describe the coupling strength of the feedback from the membrane potential and the relative timescale between the membrane potential and the recovery variable, and $I$ is the external current. The parameters $a = -0.02651$, $b = 0.006$, $c = 0.02$, and $I = 0$ are chosen to make the drive neuron exhibit relaxation oscillations.
	\par
\begin{figure*}[!ht]
		\centering
		\begin{subfigure}{0.3\textwidth}
			\centering
			\includegraphics[width=\textwidth]{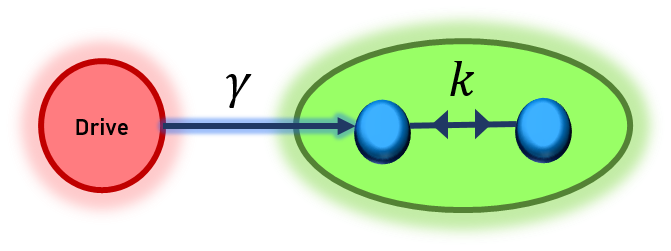}
			\caption{}
			\label{dr_2coup}
		\end{subfigure}
		\hfill
		\begin{subfigure}{0.3\textwidth}
			\centering
			\includegraphics[width=\textwidth]{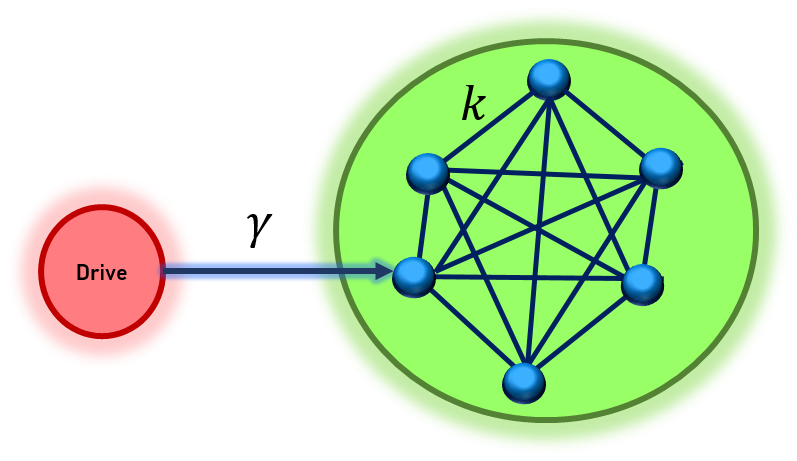}
			\caption{}
			\label{dr_Ncoup}
		\end{subfigure}
		\hfill
		\begin{subfigure}{0.3\textwidth}
			\centering
			\includegraphics[width=\textwidth]{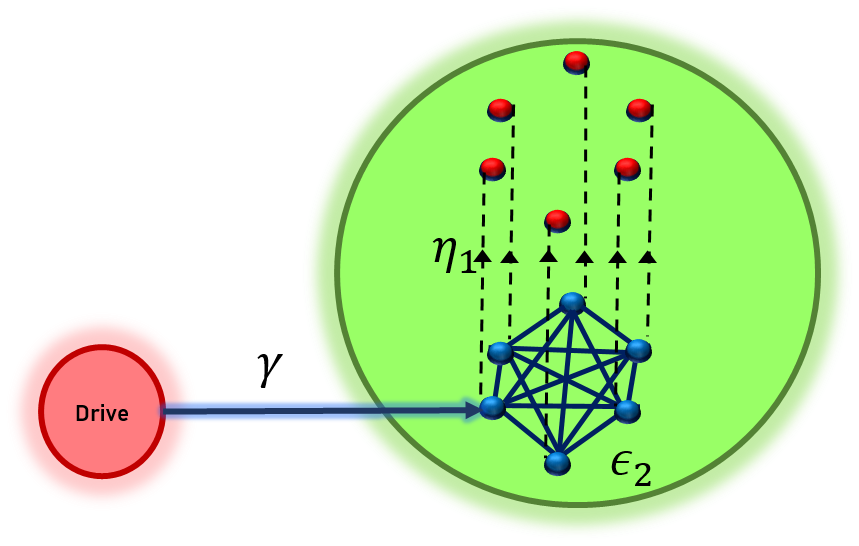}
			\caption{}
			\label{dr_2layer}
		\end{subfigure}
		\caption{(a) Drive neuron connected with one neuron of the two coupled response network, (b) drive neuron connected with a single neuron of the monolayer response network and (c) drive neuron connected to a single neuron in the Layer-2 (Blue connected nodes) in the multiplex FHN network. Layer-1 is composed of red nodes.}
		\label{dr_config}
	\end{figure*}

 The schematic diagram of the drive-response coupling configuration for the two coupled, N-coupled, and two-layer multiplex network is portrayed in Figs.~(\ref{dr_2coup}), (\ref{dr_Ncoup}) and (\ref{dr_2layer}) and the mathematical form of the three systems are given in Eqs.~(\ref{fhn_2c}), (\ref{fhn_nc}), and (\ref{multiplex}), respectively. \\
	\\
	\textbf{(i) Two coupled FHN response system}
	\begin{eqnarray}
		\dot{x}_{i}  &=& x_{i}(a_{i} - x_{i})(x_{i} - 1) - y_{i} + k\sum_{\substack{j=1 \\ i\neq j}}^{2} A_{ij}(x_{j} - x_{i}), \nonumber \\
		\dot{y}_{i}  &=& b_{i}x_{i} - c_{i}y_{i} + \gamma g(y_m, y_1), \quad i, j = 1, 2.
		\label{fhn_2c}
	\end{eqnarray}
	\textbf{(ii) N-coupled FHN response system}
	\begin{eqnarray}
		\dot{x}_{i}  &=& x_{i}(a_{i} - x_{i})(x_{i} - 1) - y_{i} + k\sum_{\substack{j=1 \\ i\neq j}}^{N} A_{ij}(x_{j} - x_{i}), \nonumber \\
		\dot{y}_{i}  &=& b_{i}x_{i} - c_{i}y_{i} + \gamma g(y_m, y_1), \quad i, j = 1, 2,...,N.
		\label{fhn_nc}
	\end{eqnarray}
	\textbf{(iii) Two-layer Multiplex FHN response Network}\\
	\begin{subequations}
		
		\textbf{Layer-1 (uncoupled neurons):}
		\begin{eqnarray}
			\dot{x}_{i,1}  &=& x_{i,1}(a - x_{i,1})(x_{i,1} - 1) - y_{i,1} +  \eta_{1}h_{1}(x_{i,1},x_{i,2}),    \nonumber \\
			\dot{y}_{i,1}  &=& b_{i}x_{i,1} - cy_{i,1},  \label{ly1} 
		\end{eqnarray}
		
		\textbf{Layer-2 (coupled neurons):}
		\begin{eqnarray}
			\dot{x}_{i,2}  &=& x_{i,2}(a - x_{i,2})(x_{i,2} - 1) - y_{i,2} \nonumber \\&& + \epsilon_{2}\sum_{\substack{j=1 \\ i\neq j}}^{N} A_{ij}(x_{j,1} - x_{i,1}) + \eta_{2}h_{2}(x_{i,1},x_{i,2}), \nonumber \\
			\dot{y}_{i,2}  &=& b_{i}x_{i,2} - cy_{i,2} + \gamma g(y_m, y_{1,2}),  \quad i = 1, 2,...,N. \label{ly2}
		\end{eqnarray}	
		\label{multiplex}
	\end{subequations}
	\par In all three cases, the parameters in the response networks are fixed such that they exhibit EEs. The Peak-over-threshold (POT) method is used to distinguish EE from normal events \cite{Review}. By this method, whenever the time evolution of the system exceeds a calculated threshold value, then the corresponding observable is said to exhibit an extreme behaviour. The threshold is calculated as 
	\begin{equation}
		x_{th} = \langle \bar{x}_{i} \rangle + n\sigma_{x}, \quad n \in \mathbb{R}\backslash\{0\}~\text{and}~n>1,
		\label{threshold}
	\end{equation}
	where $\langle \bar{x}_{i}\rangle$ is the mean of the peaks, $\sigma_{x}=\sqrt{\big(\bar{x}_{i})^2  - \langle\bar{x}_{i}\rangle^{2}}$ is the standard deviation. {\textcolor{black}{Here $n$ is a positive real number that quantifies the rareness of EE with respect to the spatial location \cite{Amritkar}. We fix $n=8$ throughout this work to estimate $x_{th}$ and $n=8$ clearly indicates that the observed events are rare enough to be qualified as EEs.}}
	\section{Mitigation analysis}
	\label{mitigation}
	\subsection{Two coupled FHN response system}
	
	\par Initially, we begin our analysis by considering two coupled FHN in the response network as given in Eq.~(\ref{fhn_2c}) where $i = 1,2$. The parameters are $a_{i} = -0.025794$, $b_{1} = 0.0065$, $b_{2} = 0.0135$, $c_{i} = 0.02$ and $k = 0.128$. $A_{ij}$ is the adjacency matrix of the coupling, which takes the value $A_{ij} = 1$ if the nodes $i$ and $j$ are connected; otherwise, $A_{ij} = 0$. The drive-response interaction is considered in the linear diffusive coupling form \cite{chimerareview}
	\begin{equation}
		g(y_m, y_1) = (y_m - y_1).
		\label{ms_int}
	\end{equation}
	\begin{figure*}[!ht]
		\centering
		\includegraphics[width=1.0\textwidth]{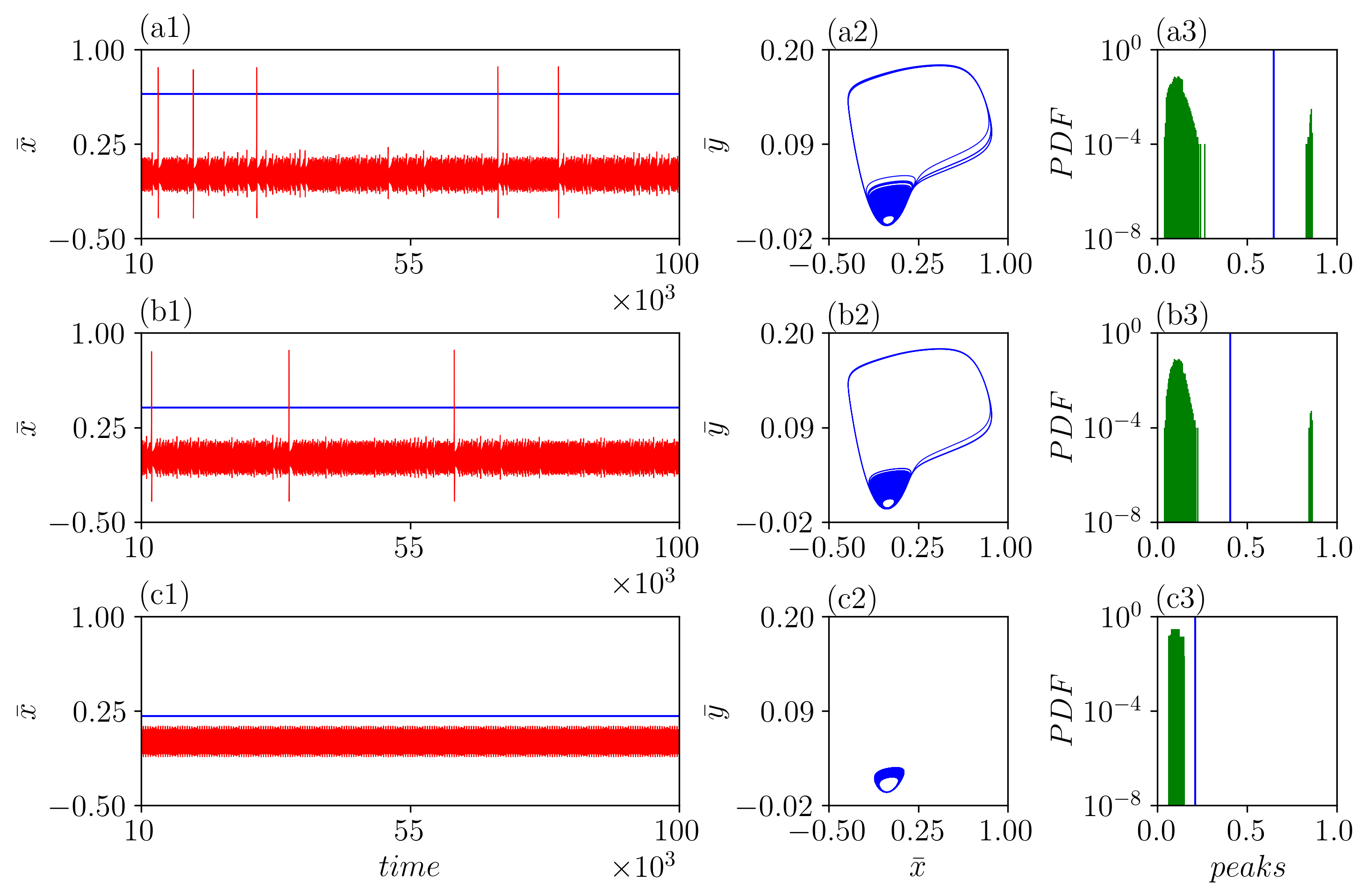}	
		\caption{The time series, phase portraits, and probability distribution plots \textcolor{black}{for two coupled systems (Eq.~\ref{fhn_2c})} are shown in the 1st, 2nd, and 3rd columns, respectively. Rows (a1-a3), (b1-b3) and (c1-c3) correspond to plots for the coupling strength $\gamma = 1.0 \times 10^{-8}$, $1.0 \times 10^{-6}$ and $1.0 \times 10^{-4}$. The blue solid line in time series and probability distribution plots corresponds to the threshold $x_{th}$.}
		\label{2cn_ts_y1}
	\end{figure*}
	\par The recovery variable $(y_m)$ of the drive neuron is coupled to the $y$-variable of the target neuron, here the first neuron $(y_1)$ is the target neuron in the response network. The control parameter $\gamma$ is the coupling strength between the drive and response neuron. The response system (\ref{fhn_2c}) exhibits EEs for the set of values given above \cite{Ansmann}. When $\gamma=0$, there is no influence of the drive on the response. Now, when the connection between the drive and the response is turned on and subsequently when $\gamma$ is increased to $1.0\times10^{-8}$, the EEs in the response network still sustain. Here, in the response network, each EE is categorized by the large amplitude peaks crossing the threshold $x_{th}$ (Eq.~(\ref{threshold})). Further, these can be visualised as a long excursion of the trajectory from the bounded chaotic region in the phase space. These results are evident from the plots in Figs.~\ref{2cn_ts_y1}(a1) and \ref{2cn_ts_y1}(a2). The probability of EE can be confirmed from the probability distribution function (PDF) plot in Fig.~\ref{2cn_ts_y1}(a3) as there is a non-zero probability beyond the threshold (blue vertical line). Upon increasing the coupling strength to $\gamma = 1.0\times10^{-6}$, the number of large amplitude peaks is reduced, but the EEs persists (Figs.~\ref{2cn_ts_y1}(b1)-\ref{2cn_ts_y1}(b2)). The probability of EE is decreased in the PDF plot as well, as observed from Fig.~\ref{2cn_ts_y1}(b3). Further increasing $\gamma$ to $1.0\times10^{-4}$ leads to the complete mitigation of EEs as no large-amplitude peaks cross the threshold (Fig.~\ref{2cn_ts_y1}(c1)) and there is no long excursion of the trajectory from the bounded chaotic region (Fig.~\ref{2cn_ts_y1}(c2)). The mitigation of EEs over the long time can be confirmed in the PDF plot in Fig.~\ref{2cn_ts_y1}(c3) as there is no probability for the occurrence of EE crossing the threshold.    
	\begin{figure}[!ht]
		\centering
		\includegraphics[width=0.5\textwidth]{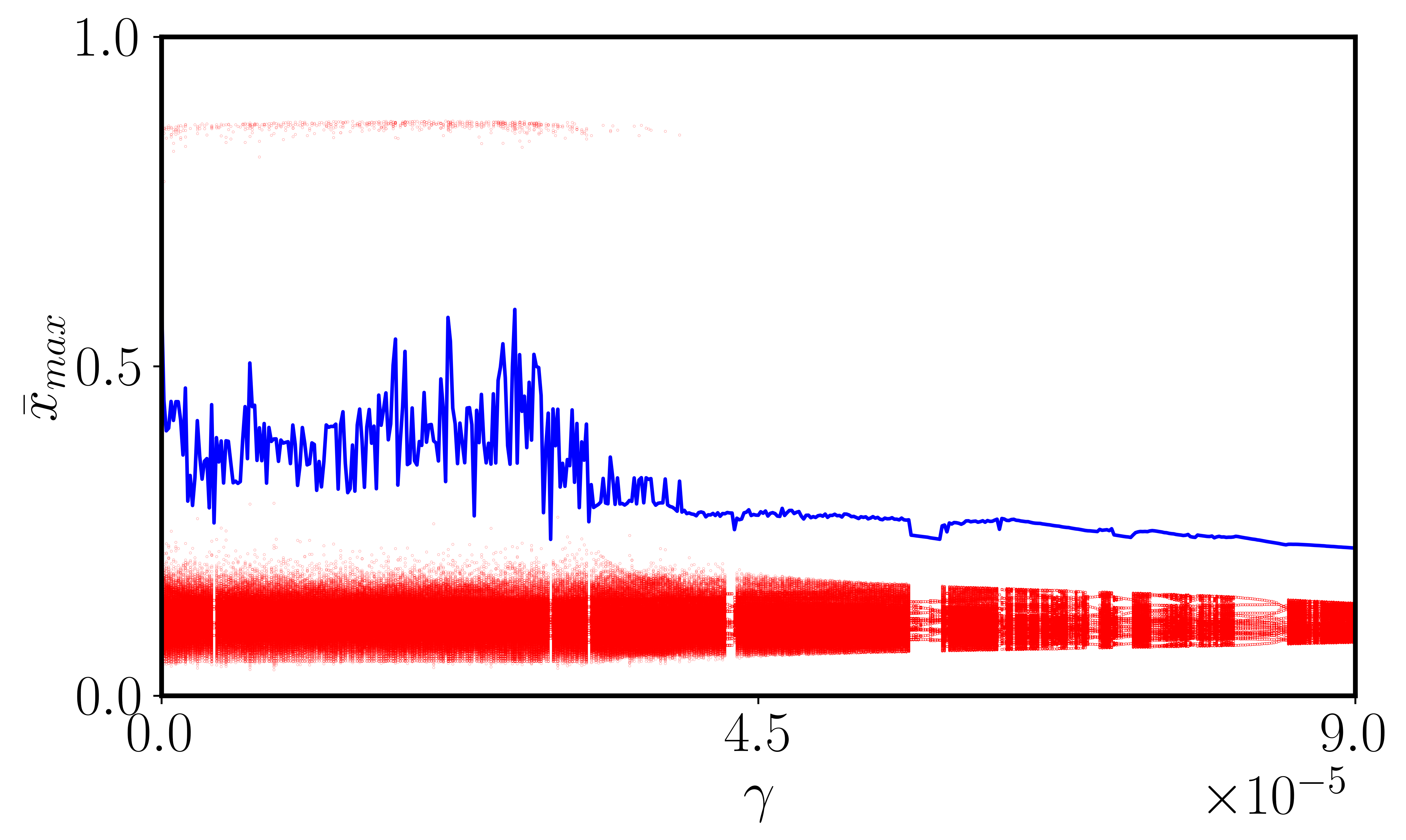}	
		\caption{Bifurcation diagram of \textcolor{black}{the two coupled systems (Eq. \ref{fhn_2c})} by collecting the peaks of $\bar{x}$ as a function of $\gamma$. The blue solid line indicates the threshold $x_{th}$.}
		\label{2cn_bif_y2}
	\end{figure}
	\par For a better perspective, the gradual mitigation of EEs is illustrated through the bifurcation diagram (see Fig.~(\ref{2cn_bif_y2})), plotted by collecting all the local maxima (peaks) of $\bar{x}_{max}$. From the plot, EEs can be visualized as the enlargement of the bounded chaotic attractor (maxima above the threshold). We can observe that EEs sustains for up to $\gamma\approx4.05\times10^{-5}$. After that, a slight increase in $\gamma\approx4.06\times10^{-5}$ leads to the sudden reduction in the size of the chaotic attractor, which indicates the mitigation of EEs. There are a few periodic windows in between the chaotic regime, which occur through saddle-node bifurcation. Further, an increase in $\gamma$ transits the dynamics from a chaotic attractor to the periodic state. For clearer visualization, we have shown only the transition of mitigation of EEs into a chaotic state. 
	\subsection{N-coupled FHN response system}
	\label{miti-mono}
	\par Next, we consider a globally coupled monolayer FHN system (as in Eq.~(\ref{fhn_nc})) in the response network. In Eq.~(\ref{fhn_nc}), $a_{i} = -0.02651$, $c_{i} = 0.02$, $k = 0.00128$ and the parameters $b_i$ are distributed from the expression $b_{i} = 0.006 + 0.008(\frac{i - 1}{N - 1})$. $A_{ij}$ is the adjacency matrix, which takes the value $A_{ij} = 1$ if the nodes $i$ and $j$ are connected; otherwise, $A_{ij} = 0$. We consider N=101 number of neurons in the response system. The drive-response interaction is the same as in the previous case, which follows Eq.~(\ref{ms_int}). In this case, also, the drive neuron is connected only to the first neuron in the response network. The influence of the drive neuron then spreads from neuron 1 to other neurons in the response network due to the global coupling topology. Throughout our analysis, the parameters of the response network are fixed in such a way that the dynamics comprises of chaotic small amplitude oscillations as well as extreme large amplitude oscillations. 
	\begin{figure*}[!ht]
		\centering
		\includegraphics[width=1.0\textwidth]{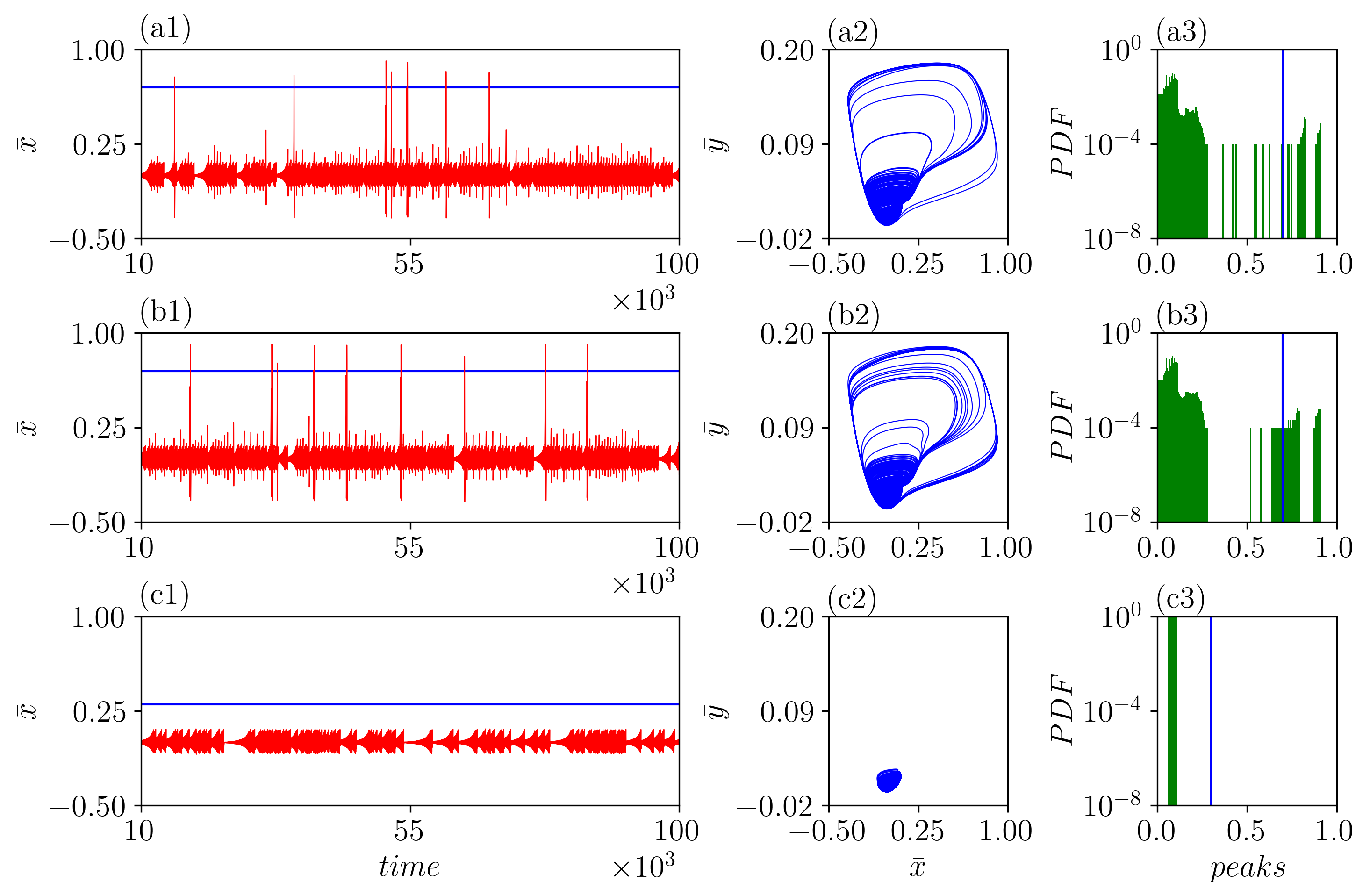}	
		\caption{The 1st, 2nd, and 3rd columns depict the time series, phase portraits, and probability distribution plots of the \textcolor{black}{ monolayer network represented by Eq.~(\ref{fhn_nc})}, respectively. Rows (a1-a3), (b1-b3) and (c1-c3) correspond to the plots for the coupling strength $\gamma = 1.0 \times 10^{-8}$, $1.0 \times 10^{-5}$ and $1.0 \times 10^{-2}$. The blue solid line in time series and probability distribution plots corresponds to the threshold $x_{th}$.}
		\label{ncn_ts_y1}
	\end{figure*}
	\par Now, we introduce the influence of the drive on the response by fixing $\gamma = 1.0\times10^{-8}$. For such low magnitudes of $\gamma$, EEs still exist in the system. This can be visualized through the large amplitude peaks crossing the threshold in Fig.~\ref{ncn_ts_y1}(a1) and through the long excursion of the trajectory from the bounded chaotic region in the phase space in Fig.~\ref{ncn_ts_y1}(a2). The corresponding PDF plotted in Fig.~\ref{ncn_ts_y1}(a3) confirms the existence of non-zero probability beyond the threshold, even for large times. Next, increasing the coupling strength ($\gamma$) to $1.0\times10^{-5}$ we can visualise no impact and the EE still persists (Figs.~\ref{ncn_ts_y1}(b1-b2)), whereas the probability of occurence of EE is decreased, as observed from the PDF plot in Fig.~\ref{ncn_ts_y1}(b3). Finally, increasing $\gamma$ to $1.0\times10^{-2}$ leads to the complete mitigation of EEs since no large amplitude peaks cross the threshold (Fig.~\ref{ncn_ts_y1}(c1)) and no long excursion of the trajectory appears from the bounded chaotic region (Fig.~\ref{ncn_ts_y1}(c2)). The probability for occurence of EE behind the threshold in Fig.~\ref{ncn_ts_y1}(c3) and the mitigation of EEs, even for long times, can be confirmed from the PDF plot.
	\begin{figure}[!ht]
		\centering
		\includegraphics[width=0.5\textwidth]{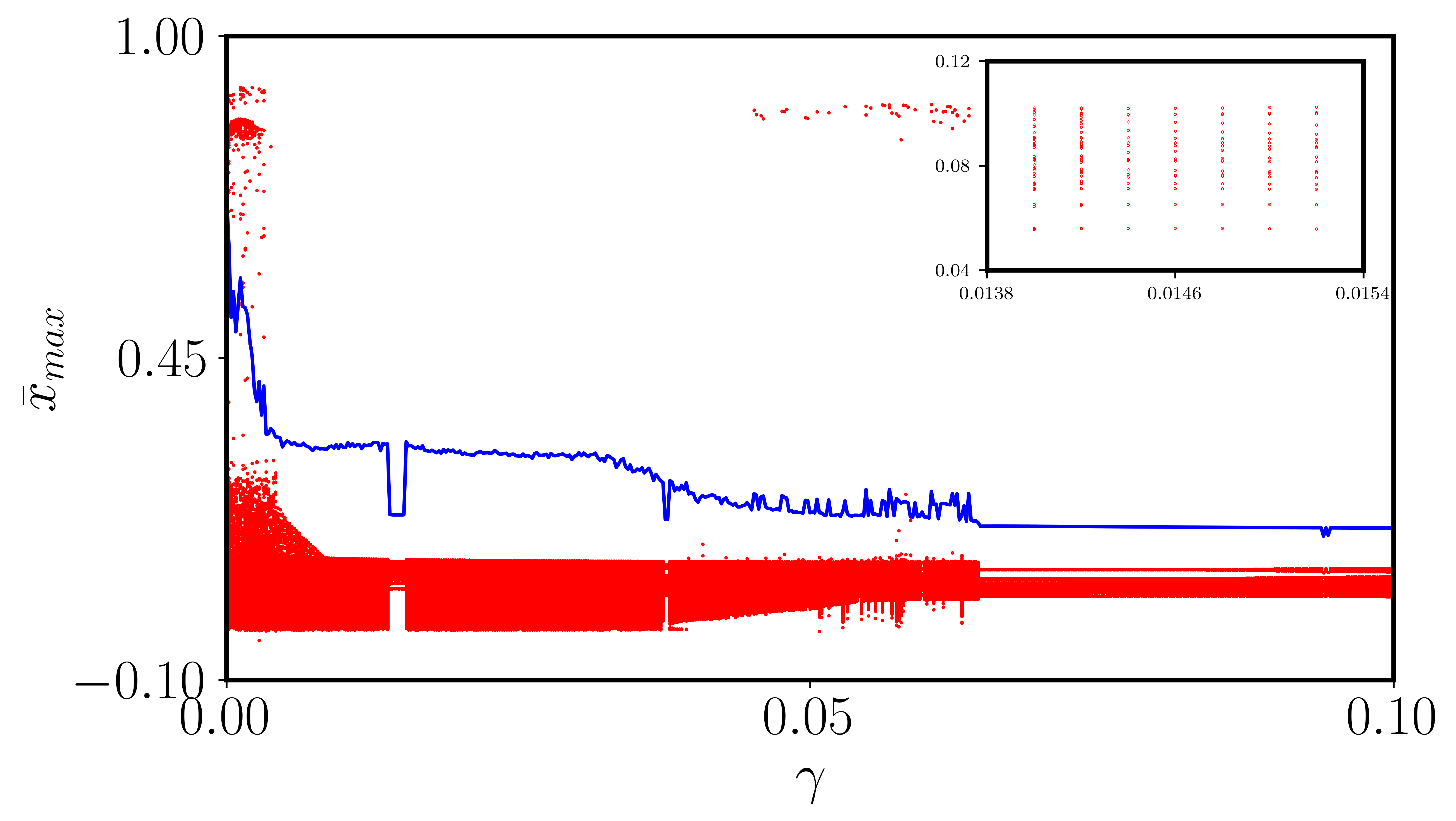}	
		\caption{Bifurcation diagram of \textcolor{black}{the monolayer network (Eq. \ref{fhn_nc})} with varying the coupling strength $\gamma$ as the control parameter. The blue solid line indicates the threshold $x_{th}$.}
		\label{ncn_bif_y2}
	\end{figure}
	\linebreak
	\par The bifurcation diagram in Fig.~(\ref{ncn_bif_y2}) shows the mitigation of EEs in $\bar{x}_{max}$ for the gradual increase in the value of the drive-response coupling strength $\gamma$. The coupling parameter ($\gamma$) has no impact on EEs till $\gamma\approx4.06\times10^{-3}$, and a slight increase in $\gamma\approx4.07\times10^{-3}$ leads to the mitigation of EEs. The chaotic attractor is stable upto $\gamma\approx4.43\times10^{-2}$ and again EEs occurs for short range ($\gamma\approx4.44\times10^{-2} - 6.24\times10^{-2}$). Further, an increase in $\gamma$ completely mitigates EEs, and only the chaotic state persists. Through saddle-node bifurcation, a few periodic windows occur in between the chaotic regimes.
	\subsection{Two-layer multiplex FHN response system}
	\par Next, we extend our analysis by considering two-layer multiplex FHN system in the response network, as given in Eqs. (\ref{ly1}) and (\ref{ly2}). Each layer in the network has the same number of neurons ($N=101$) and the neurons in layer-1 are connected only to their replicas in layer-2. The neurons inside layer-1 are uncoupled, whereas the neurons within layer-2 are globally coupled. We fix the values of the parameters as $a = -0.02651$, $c = 0.02$ while the parameters $b_{i}$ are distributed from the expression $b_{i} = 0.006 + 0.008(\frac{i - 1}{N - 1})$. The parameter $\epsilon_{2}$ is the intralayer coupling strength of layer-2, whereas $\eta_{1}$ and $\eta_{2}$ are the interlayer coupling strengths from layer 2 to 1 and from layer 1 to 2, respectively. When $\eta_{1} = \eta_{2} = 0.0$ (no transfer of information between the layers), layer-1 exhibits chaotic dynamics and layer-2 exhibits EE. For the analysis, we choose $\eta_{1} = 1.0$ and $\eta_{2} = 0.0$, which corresponds to a two-layer network with unidirectional interaction (layer-2 to layer-1) where both layers exhibit EE \cite{Shashang2}. $A_{ij}$ is the adjacency matrix of the layer-2 intralayer coupling which takes the value $A_{ij} = 1$ if the nodes $i$ and $j$ are connected otherwise, $A_{ij} = 0$. The values of the parameters in layer-1 are the same as those in layer-2. The terms $h_{1}(x_{i,1},x_{i,2})$ and $h_{2}(x_{i,1},x_{i,2})$ are given by $x_{i,2} - x_{i,1}$ and $x_{i,1} - x_{i,2}$ respectively. We couple the drive neuron to the first neuron in layer-2, and the corresponding drive-response interaction is given by 
	\begin{equation}
		g(y_m, y_{1,2}) = (y_m - y_{1,2}).
		\label{ms_int_ly}
	\end{equation}
	\begin{figure*}[!ht]
		\centering
		\includegraphics[width=1.0\textwidth]{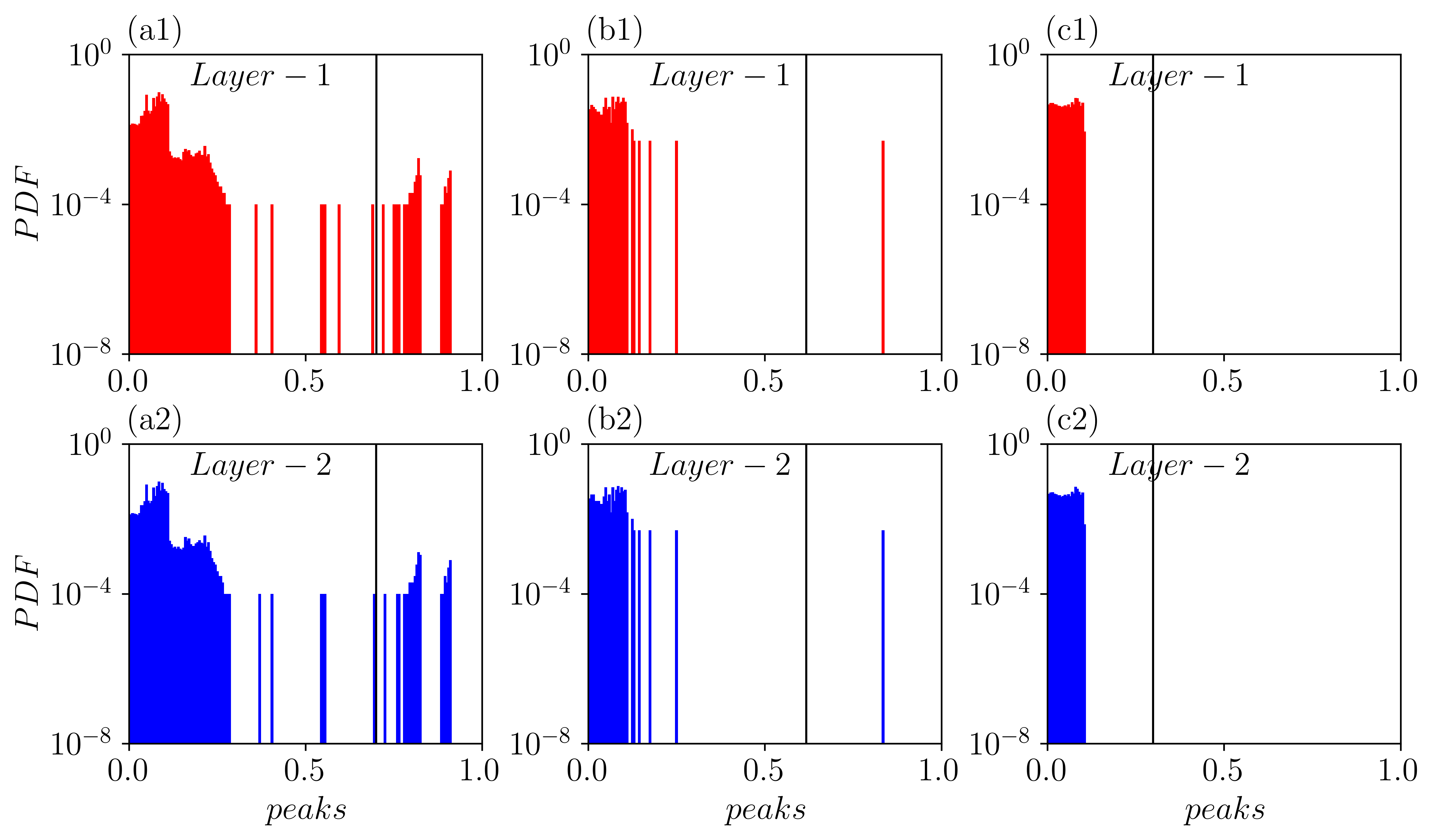}	
		\caption{The mitigation of EEs in \textcolor{black}{two-layer multiplex network (Eq. \ref{multiplex})} is confirmed through the probability distribution plots. The top and bottom rows correspond to layer-1 and layer-2 and Columns 1-3 denote the plots for the parameter values $\gamma = 1.0 \times 10^{-8}$, $1.0 \times 10^{-5}$, and $1.0 \times 10^{-2}$. The black solid vertical line indicates the threshold $x_{th}$.}
		\label{fhn_ly1ly2_pdf}
	\end{figure*}
	\par We start our analysis with $\gamma = 1.0\times10^{-8}$ where the EE in the system sustain in both layers. To avoid redundancy, we show our results only with the PDF plots. In this connection, the probability of EE over a long time can be confirmed from the non-zero probability after the threshold in the Figs.~\ref{fhn_ly1ly2_pdf}(a1) and \ref{fhn_ly1ly2_pdf}(a2). Increasing $\gamma$ to $1.0\times10^{-5}$ decreases the probability appreciably, but EEs persist (Figs.~\ref{fhn_ly1ly2_pdf}(b1)-\ref{fhn_ly1ly2_pdf}(b2)). Finally, increasing $\gamma$ to $1.0\times10^{-2}$ leads to the mitigation of EEs as there is no probability behind the threshold (Figs.~\ref{fhn_ly1ly2_pdf}(c1)-\ref{fhn_ly1ly2_pdf}(c2)). We have also carried out the bifurcation analysis for both the layers to observe how EEs are mitigated and found that both layers follow the same mitigation pattern as in the monolayer case discussed in the sub-section \ref{miti-mono}. Mitigation happens even in uncoupled layer-1, because the unidirectional coupling from layer-2 to layer-1 promotes the same dynamics between the layers. So, EEs get controlled in both layers. 
	\section{Mitigation Mechanism}
	\label{mecha}
	Imperfect phase synchronization and excitation of protoevents are the potential precursors \cite{Ansmann}, behind the EE that are present in the response network particularly when it is two-coupled and monolayer, respectively. In this section, we identify the mechanism/route behind the suppression that took place in the response network. 
	\subsection{Phase Lock Breaking}
\begin{figure*}[!ht]
		\centering
		\includegraphics[width=0.85\textwidth]{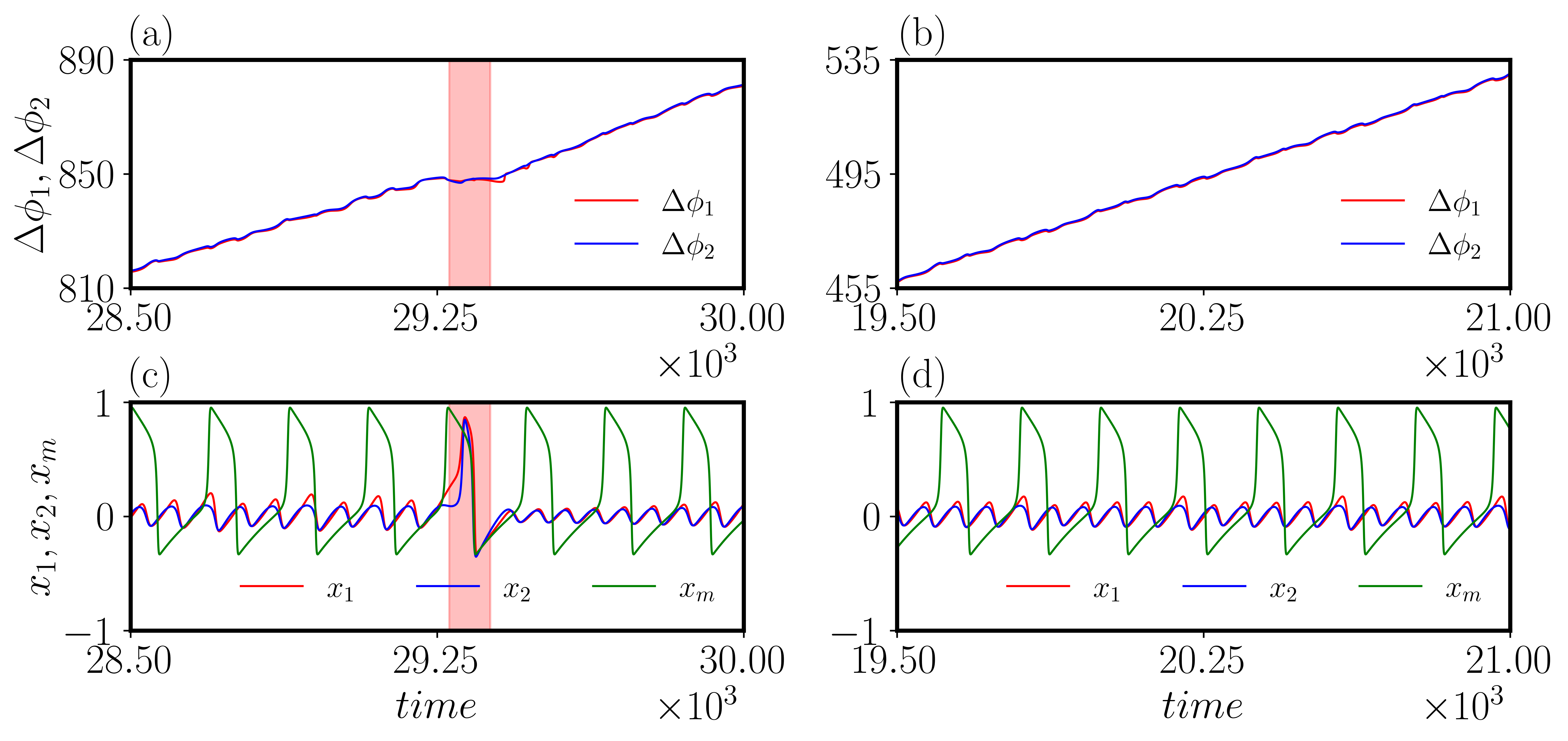}
		\caption{(a, b) Variation of phase differences ($\Delta\phi_{1}$ and $\Delta\phi_{2}$) with respect to time and (c, d) time series of drive ($x_{m}$) and response neurons ($x_{1}$ and $x_{2}$) for the coupling parameters $\gamma=1.0\times10^{-8}$ and $\gamma=1.0\times10^{-4}$ \textcolor{black}{of the two coupled systems (Eq.~(\ref{fhn_2c}))}. The highlighted portion indicates the EE zone where phase locking occurs.}
		\label{fhn_2c_ps_final}	
	\end{figure*}

	We start our investigation through the phase analysis, where we calculate the instantaneous phase ($\phi$) of drive ($\phi_{m}$) and response neurons ($\phi_{1}$ and $\phi_{2}$) using a Hilbert transform technique \cite{Wasif}. Then we estimate the phase difference between the drive and response neurons, which is $\Delta\phi_{1}=\phi_{m}-\phi_{1}$ and $\Delta\phi_{2}=\phi_{m}-\phi_{2}$. We plot the result in the form of time evolution of phase differences and individual neurons in Fig.~\ref{fhn_2c_ps_final} for two different parameters ($\gamma=1.0\times10^{-8}$ and $\gamma=1.0\times10^{-4}$). We observe from Fig.~\ref{fhn_2c_ps_final}(a) that the phase difference increases linearly and stays constant for some time (highlighted portion) and then again starts to increase linearly. The portion where the phase difference remains constant indicates that the two response neurons are phase locked in that region with the drive neuron. This phase locking occurs exactly at the time when EEs emerge in the response network (see the corresponding highlighted portion in Fig.~\ref{fhn_2c_ps_final}(c)). The same pattern of phase locking occurs everywhere in time whenever an EE emerges. This phase locking persists for all the parameters of the coupling strength wherever EE occur. In such a scenario, when the coupling strength between the drive and the response is increased, the drive neuron's influence on the response becomes higher and particularly at $\gamma = 1.0\times10^{-4}$, the system gains sufficient coupling strength to break the phase locking occurring in the response network. As this phase locking is broken, the EEs in the response network are mitigated. Thus, there is no region where the phase difference is constant, and the phase differences increase linearly through time. These results can be observed from the plots given in Figs.~\ref{fhn_2c_ps_final}(b) and \ref{fhn_2c_ps_final}(d), where there is no regime of constant phase difference and EEs.
	\subsection{Disruption of protoevents frequency}
	\subsubsection{\textbf{N-coupled FHN network}}
	\par As mentioned previously, protoevents are responsible for the emergence of EE in the response network when the network has a monolayer globally coupled architecture \cite{Ansmann}. Generally, protoevents denote the simultaneous excitation of 22 or more units ($x_{i} > 0.6$). Occurrence of protoevents further stimulates all the oscillators in the network to excite synchronously, producing the EE in the average membrane potential.
	\begin{figure*}[!ht]
		\centering
		\includegraphics[width=1.0\textwidth]{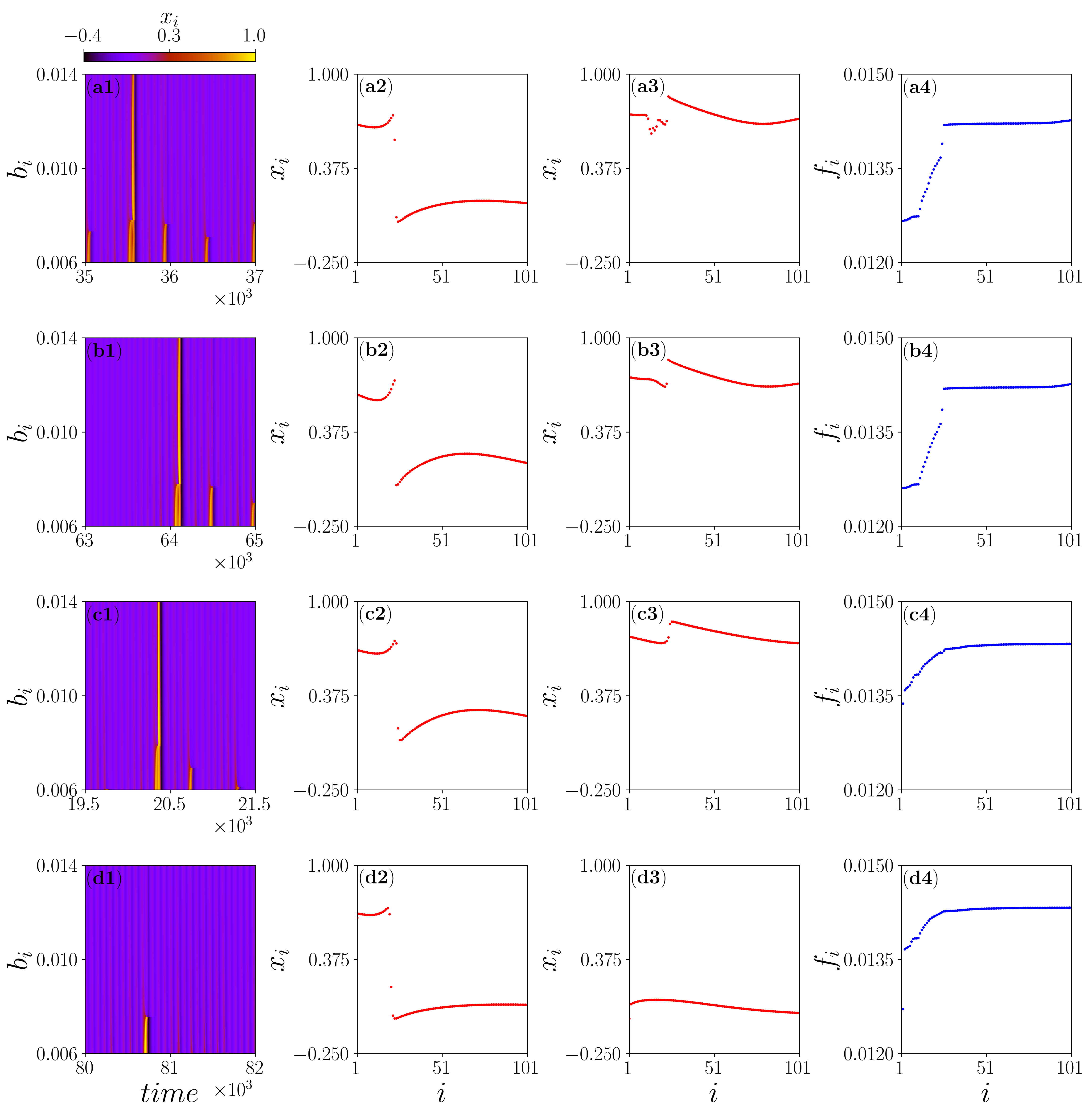}
		\caption{The spatiotemporal, snapshots at $t = 35400, 35600$, and frequency plots of \textcolor{black}{the monolayer network} are displayed in the 1st, 2nd, 3rd and 4th columns. While the rows denotes the plots for $\gamma$ = $1.0\times10^{-8}$, $1.0\times10^{-5}$, $1.0\times10^{-3}$ and $4.07\times10^{-3}$, respectively.}	
		\label{ncn_sssf}
	\end{figure*}
	\par We start our analysis with the help of a spatiotemporal plot, snapshots and frequency plots. Columns 1-4 of Fig.~\ref{ncn_sssf} represent, respectively, the spatiotemporal plot, snapshots, and frequency plot for various values of $\gamma$. Columns 2 and 3 represent the snapshots for protoevents and EE, respectively. When $\gamma=1.0\times10^{-8}$, the system exhibits EE between $t\approx35400$ and $35600$ which can be observed from the spatiotemporal plot in Fig.~\ref{ncn_sssf}(a1). The emergence of protoevents and EE can be confirmed from the snapshots in plots Fig.~\ref{ncn_sssf}(a2), where 23 units are excited ($>0.6$), and Fig.~\ref{ncn_sssf}(a3), where all 101 units are excited, exhibiting EE. Fig.~\ref{ncn_sssf}(a4) represents the frequency of the oscillators during an EE, which is split into two clusters, where the first 23 oscillators (protoevents) are in a hierarchical frequency, while the remaining oscillators have the same frequency. A similar pattern is observed in all four plots when the coupling strength, $\gamma=1.0\times10^{-5}$ (Figs.~\ref{ncn_sssf}(b1)-\ref{ncn_sssf}(b4)). Next, when the coupling strength is $\gamma=1.0\times10^{-3}$, EE sustains, and the spatiotemporal and the snapshot resemble the same as it was for the previous coupling values (Figs.~\ref{ncn_sssf}(c1)-\ref{ncn_sssf}(c3)). In contrast with the previous values of $\gamma$, for the present value, we can observe a change of pattern in frequency of the oscillators involved in the protoevents as shown in Fig.~\ref{ncn_sssf}(c4). The response neuron, which is attached to the drive, gets detached and oscillates solitarily at a different frequency. 
	Finally, when $\gamma=4.07\times10^{-3}$, there is a sudden mitigation of EE (Fig.~\ref{ncn_sssf}(d1)), because less than 22 oscillators excite. The plots in Figs.~\ref{ncn_sssf}(d2)-\ref{ncn_sssf}(d3) correspond to protoevent and non-EE zones. In this case, we can also observe that the single neuron connected to the drive fires solitarily at a different frequency.
	\par The critical observation here is that, as the coupling strength is increased, the drive neuron first affects the frequency pattern of the protoevent oscillators, which are responsible for the emergence of EE, and also detaches the response neuron from the protoevent oscillators and makes it to oscillate at a separate frequency. Upon increasing the coupling strength further, the frequency of the response neuron decreases.
	\subsubsection{\textbf{Two-layer FHN network}}
	\begin{figure*}[!ht]
		\centering
		\includegraphics[width=1.0\textwidth]{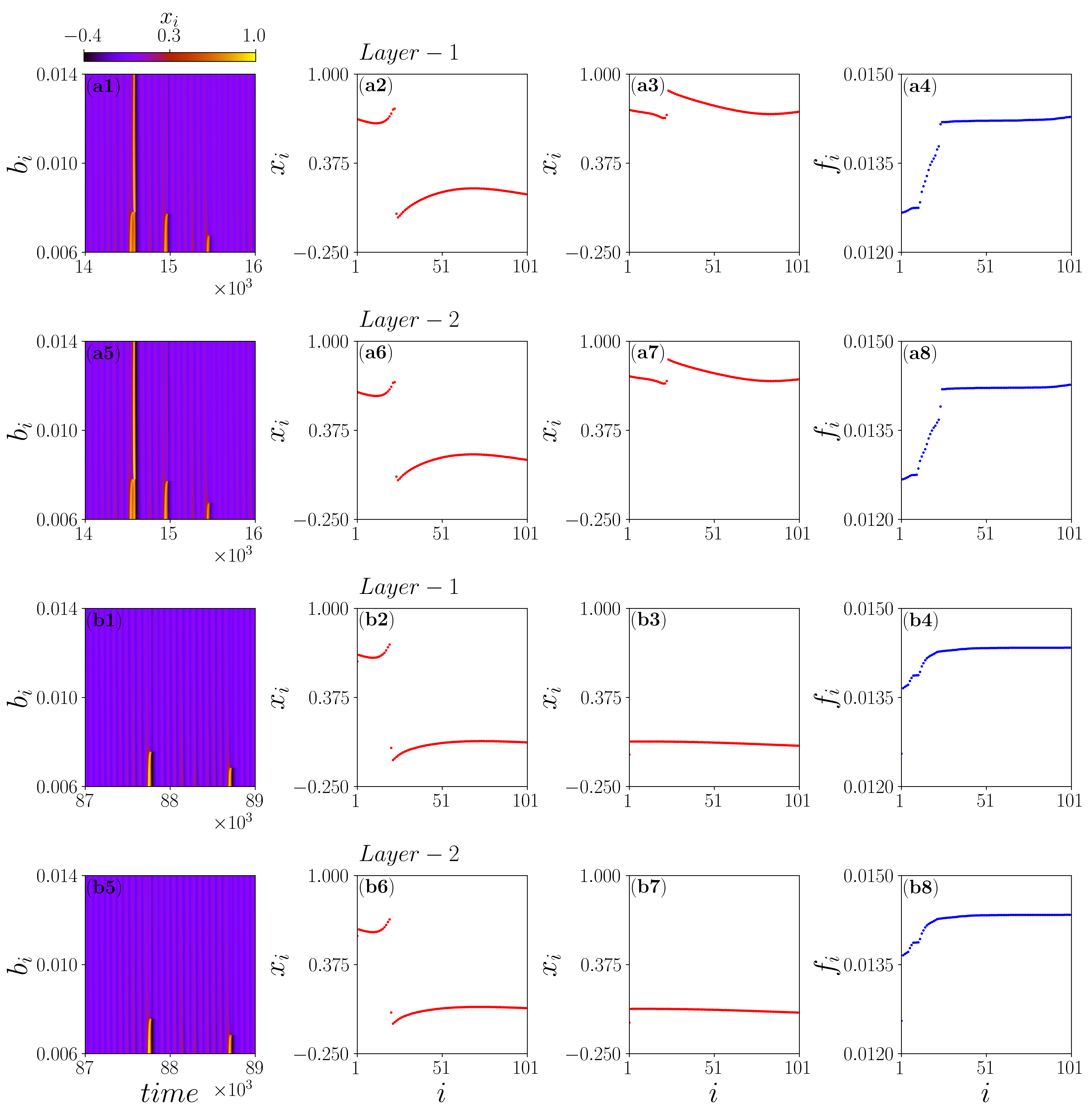}
		\caption{The spatiotemporal, snapshots at $t = 35400, 35600$, and frequency plots of \textcolor{black}{the two-layer multiplex network} are displayed in the 1st, 2nd, 3rd and 4th columns. While the rows one (layer-1) and two (layer-2) denote the plots for $\gamma$ = $1.0\times10^{-8}$ whereas rows three (layer-1) and four (layer-2) denote the plots for $\gamma=5.0\times10^{-3}$.}
		\label{ly_sssf_y1}	
	\end{figure*}
	\par Here we analyze the mechanism behind the mitigation of EE when the response network is a two-layered FHN multiplex network. As in the previous case, here also, we give the spatiotemporal plot, snapshots, and frequency plots of the response network in columns 1-4 of Fig.~\ref{ly_sssf_y1}. Columns 2 and 3 represent the snapshots for protoevents and EE of the response network. The plots in the first two rows and the last two rows in Fig.~\ref{ly_sssf_y1} are the plots corresponding to $\gamma=1.0\times10^{-8}$ and $\gamma=5.0\times10^{-3}$, respectively. Since the layers are unidirectionally coupled, EE propagate from layer-2 to layer-1, and the neurons in the uncoupled layer also exhibit EE \cite{Shashang2}. After turning on $\gamma$, for very small value of $\gamma=1.0\times10^{-8}$, EE persists in both the layers (Figs.~\ref{ly_sssf_y1}(a1) and \ref{ly_sssf_y1}(a5)). The snapshots for the protoevents and EE in both the layers are shown in (Figs.~\ref{ly_sssf_y1}(a2) and \ref{ly_sssf_y1}(a6)) and (Figs.~\ref{ly_sssf_y1}(a3) and \ref{ly_sssf_y1}(a7)). The frequency of both layers is shown in Figs.~\ref{ly_sssf_y1}(a4) and \ref{ly_sssf_y1}(a8). Upon increasing $\gamma$ systematically, at $\gamma=5.0\times10^{-3}$, there is an abrupt mitigation of EE. Which means, only less than 22 neurons participate in the synchronous protoevent process (Figs.~\ref{ly_sssf_y1}(b1)-\ref{ly_sssf_y1}(b3) and 
	Figs.~\ref{ly_sssf_y1}(b5)-\ref{ly_sssf_y1}(b7)). The frequency of the oscillator participating in the protoevent is altered in such a way that the solitary response neuron connected with the drive and its replica in the uncoupled layer oscillate at the same frequency (Figs.~\ref{ly_sssf_y1}(b4) and \ref{ly_sssf_y1}(b8)), detaching from the rest of the oscillators. As in the previous case, in this case also, EE in both layers are mitigated due to the impact of the drive neuron on the frequency pattern of the protoevent oscillators.
	\par {\textcolor{black}{Next we have carried out the synchronization analysis to observe whether the interlayer synchronization has any influence on the mitigation of EEs. To measure the synchronization between nodes of a layer with the replicator nodes, we define interlayer synchronization error as, \cite{Shashang2}}
		\textcolor{black}{\begin{equation}
				S^{inter}_{E} = \displaystyle \lim_{T\to \infty} \frac{1}{T} \int_0^ T \sum_{j=1}^{N}\frac{\lvert \lvert \mathbf{x}_{j,2}(t) - \mathbf{x}_{j,1}(t) \rvert\rvert}{N} dt.
				\label{se_eqn}
	\end{equation}}}
	\par
\begin{figure}[!ht]
		\centering
		\includegraphics[width=1.0\linewidth]{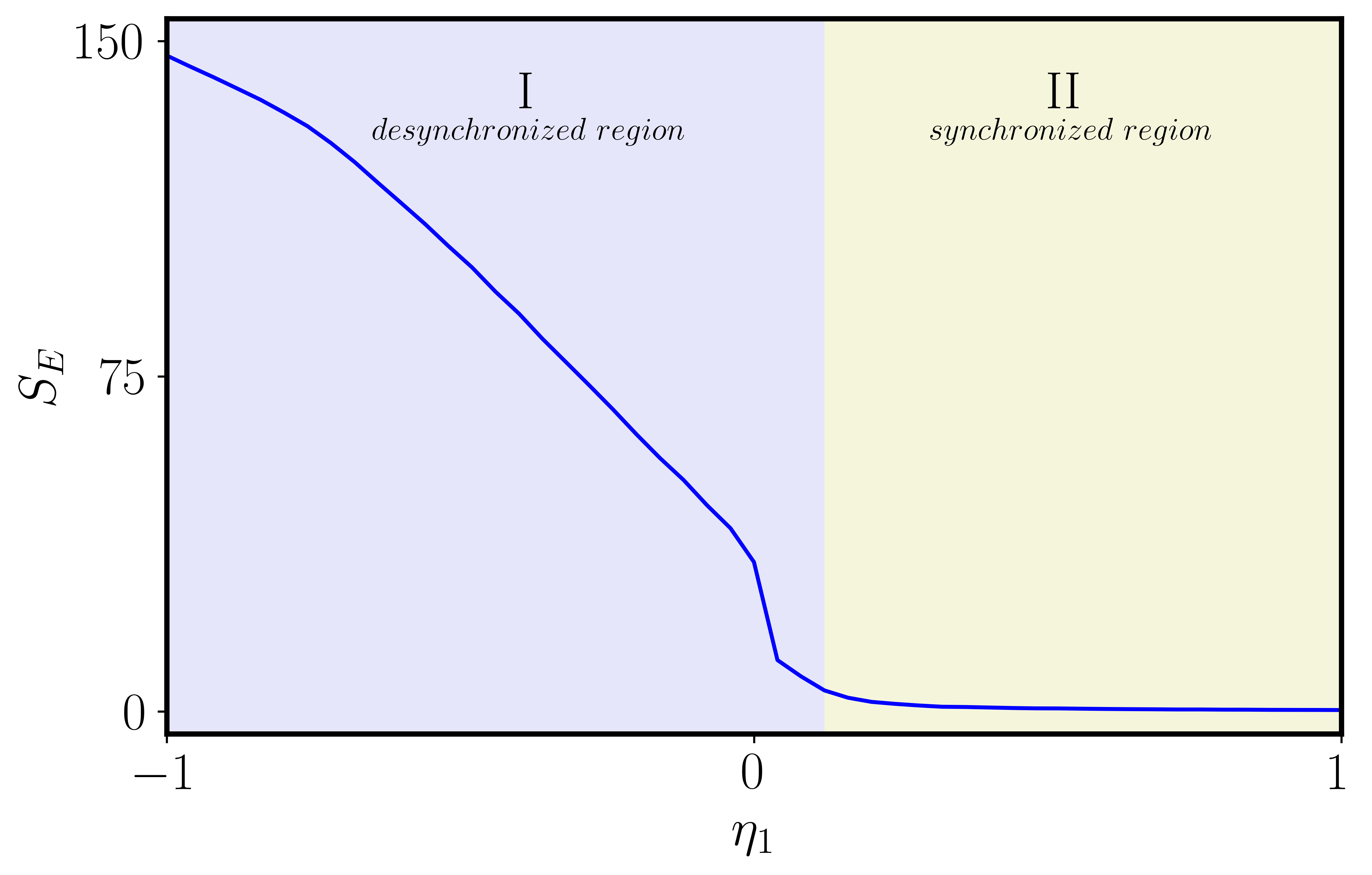}	
		\caption{Variation of interlayer synchronization error ($S_{E}$) with interlayer coupling ($\eta_1$) as a control parameter.}
		\label{interlayer_sync_err}
	\end{figure}

 {{\color{black} Further from the perspective of synchronization, as far as the present case is considered, upon increasing the interlayer coupling strength $\eta_1$ from -1 to +1, collectively, the oscillators between the layers transit from a desynchronized state to a synchronized state as evidenced from Fig.~\ref{interlayer_sync_err}. The measure of interlayer synchronization error \cite{Shashang2} is calculated from Eq. (\ref{se_eqn}). Specifically, in the inhibitory regions ($\eta_1<0$), the oscillators are desynchronized while in the excitatory regions ($\eta_1>0$), the oscillators are  synchronized. In a recent work \cite{Shashang2}, we have shown that EEs can propagate across layers in both the excitatory and inhibitory interlayer couplings, regardless of whether synchronization  occurs or not. In the present case, when the drive neuron is activated, the propagation of the control is also found to occur irrespective of whether the oscillators between the layers are synchronized or not. This ascertains the fact that the propagation of the control is mediated by the interlayer coupling strength itself, despite synchronization.}}
	\section{\textbf{$\gamma$}-parameter shift in controlling EE}
	\label{pshift}
	\begin{figure}[!ht]
		\centering
		\includegraphics[width=0.5\textwidth]{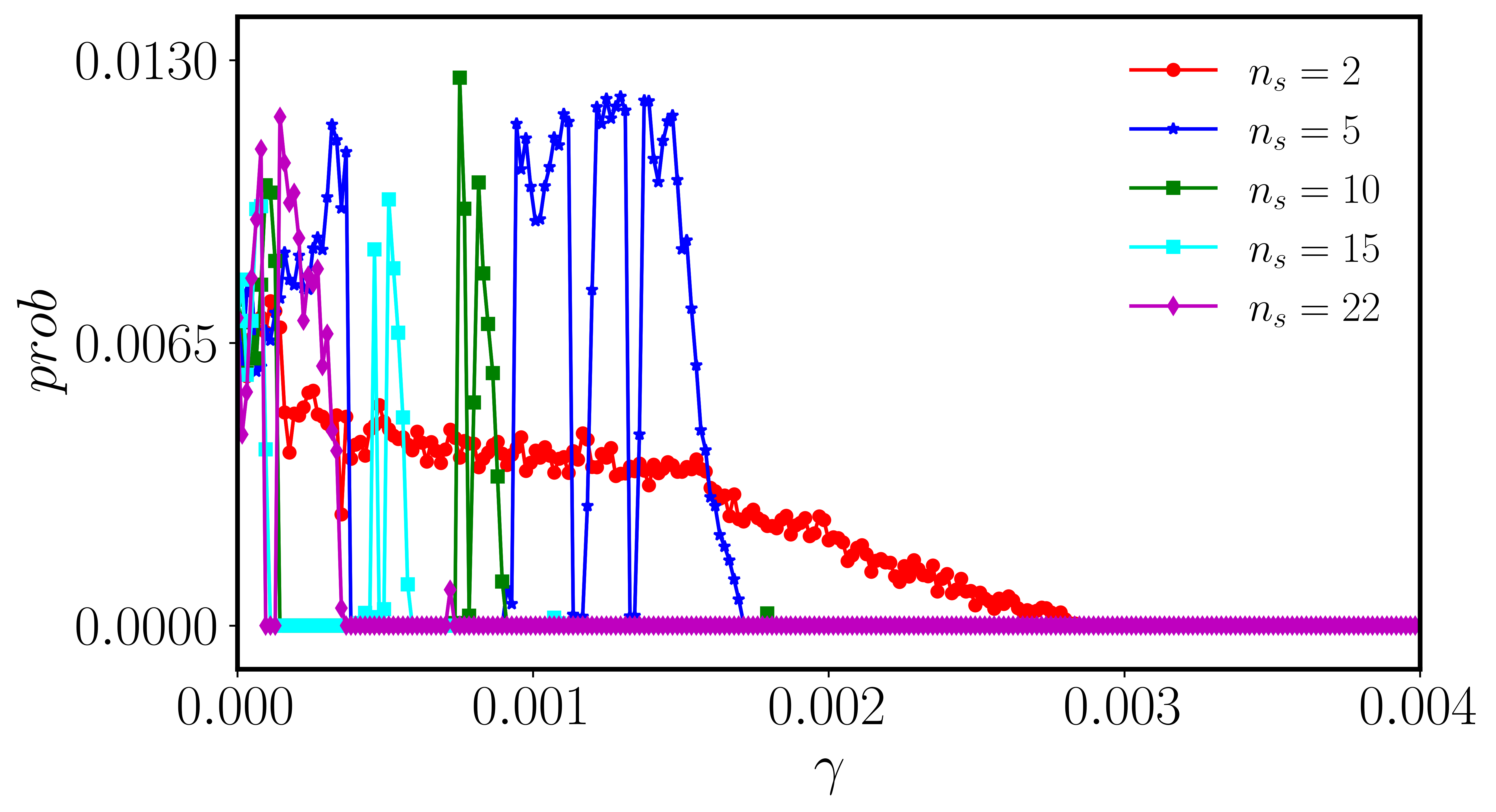}
		\caption{Variation of probability of EE as a function of $\gamma$. The probability is calculated for various numbers of response neurons ($n_s$) connected to the drive neuron.}	
		\label{fhn_ncn_prob3}
	\end{figure}
	\par So far, we have investigated the mitigation of EEs in the response network when the drive neuron is connected only to one neuron in the response network. In this section, we extend this study by increasing the number of neurons in the response network that are connected to the drive neuron and determine the relation between the number of response neurons and the mitigation of EE. For this purpose, we calculate the probability of occurrence for EE by varying $\gamma$ and the results are depicted in Fig.~\ref{fhn_ncn_prob3}. Initially, the drive neuron is connected only to one neuron in the N-coupled system, and when $\gamma\approx4.07\times10^{-3}$, the EEs are mitigated (see Fig.~\ref{ncn_bif_y2}). From Fig.~\ref{fhn_ncn_prob3}, we can observe that when the number of response neurons is 2 (red curve in Fig.~\ref{fhn_ncn_prob3}), the mitigation of EEs occur at $\gamma\approx2.84\times10^{-3}$ where the probability goes to zero. Here, the mitigation occurs earlier compared to the previous case (connection to one response neuron). By increasing this count of response neurons to 5, 10, 15, 22, (blue, green, cyan and magenta curves in Fig.~\ref{fhn_ncn_prob3}), we observe that there is a backwards shift or preponement in the $\gamma$ parameter where the probability of EE transits to zero. This shift in the $\gamma$-parameter indicates that when the number of response neurons is larger, the less is the coupling strength needed to mitigate the EEs in the system. This is true in the case of a two-layer response network. The mechanism is still the same as that of the single neuron connection case, where the frequency pattern of the protoevent oscillators changes, and the response neuron detaches from the hierarchical frequency pattern. The only difference is that detachment occurs for all neurons in the response network that are connected to the drive. For example, in the case of $n_s=2$, two oscillators detach, for $n_s=5$, five oscillators detach, for $n_s=10$, ten oscillators detach and so on.
	\section{Conclusion}
	\label{conc}
	\par In this study, we have utilized the role of the drive-response coupling for the suppression of EEs in the response network. We have considered three types of coupled networks, namely two coupled, monolayer $N$-coupled, and two-layered FHN multiplex systems. The mitigation is confirmed through time series, phase portrait, and PDF plots. We find that when the response network is (i) two coupled, breaking of phase-locking mitigates the EEs and (ii) a monolayer $N$-coupled, and two-layered multiplex network, the drive neuron impacts the frequency of the protoevent oscillators. Further, in all three response networks, we found that upon increasing the coupling strength between drive and response neuron, the frequency of large amplitude peaks decreases gradually, and there is a sudden suppression of EE. We have also found that one drive neuron is sufficient to control and mitigate the EE. Increasing the number of neurons in the response connected to the drive, prepones the onset of mitigation in the coupling parameter space. {\textcolor{black}{As of now, this control mechanism is FHN system-specific and we intend to carry out the same analysis in other classes of neuron models and chaotic systems to confirm the wider application of this approach. Devising variants to generalize this observed control mechanism globally to at least a class of systems is an intriguing quest in the immediate future. Further, in the present study we have restricted our attention only on global networks. Investigating the emergence of extreme events in other static and dynamic network topologies and strategising mitigation methods in them is also an immediate problem that one can work on.}}
	
	\par The essence of the results obtained in the present work can also be understood in the following way. If a network/layer/region of neurons in the brain (in the realistic case) displays epileptic behaviour, then the brain may restrain the network from exhibiting such an abnormal behaviour by activating a coupling between the network and a single neighbour neuron that exhibits regular firing. Further, in a pragmatic sense, this work will serve as a starting point for the researchers to implement and test the practicality of the present approach in a two-step process: (i) testing the approach on a real-time epileptic data set using time series analysis and obtaining the information regarding the crucial parameters of mitigation and (ii) with the help of the obtained information, investigating the possibility of epilepsy control by externally stimulating the least number of neurons in the scalp without damage. Devising such methods is very important to treat patients with epilepsy who are allergic to chemical dosages. Moreover, it can also be tested and extended to treat other neurological disorders such as  Alzheimer’s disease, Parkinson's disease, multiple sclerosis, and Schizophrenia. These are some notable long-term potential applications of this study, and developing additional tools to mitigate these EE remains an open problem. 
	\section*{acknowledgments}
	R.S. thanks Bharathidasan University, Tamil Nadu, India, for providing the University Research Fellowship (URF). S.S. thanks SRM TRP Engineering College, India, for their financial support, vide number SRM/TRP/RI/005. The work of M.S. forms a part of a research project sponsored by the Council of Scientific and Industrial Research (CSIR) under Grant No. 03/1482/2023/23/EMR-II.
	\section*{DATA AVAILABILITY}
	The data that support the findings of this study are available within the article.
	\section*{Author Declarations}
	\subsection*{Conflict of Interest}
	The authors have no conflicts to disclose.
	\section*{Author Contributions}
	{\bf R. Shashangan}: Conceptualization (equal); Data curation (equal); Formal analysis (equal); Investigation (equal); Methodology
	(equal); Software (equal); Validation (equal); Visualization (equal); Writing – original draft (equal). {\bf S. Sudharsan}: Conceptualization (equal); Data curation (equal); Formal analysis (equal); Investigation (equal); Methodology
	(equal); Software (equal); Validation (equal); Visualization (equal); Writing – original draft (equal). {\bf Dibakar Ghosh}: Conceptualization (equal); Supervision (equal); Visualization
	(equal); Writing – review and editing (equal).
	{\bf M. Senthilvelan}: Conceptualization (equal); Supervision (equal); Visualization
	(equal); Writing – review and editing (equal).  
	\section*{Refereneces}
	
\end{document}